\documentclass[11pt]{article}
\usepackage[T1]{fontenc}
\usepackage{lmodern,amsmath,amssymb,amsthm,mathtools}
\usepackage[margin=0.95in]{geometry}
\usepackage{enumitem,microtype,needspace,booktabs}
\usepackage{tikz}
\usetikzlibrary{arrows.meta,positioning,fit}
\usepackage[hidelinks]{hyperref}
\newtheorem{theorem}{Theorem}[section]
\newtheorem{lemma}[theorem]{Lemma}
\newtheorem{corollary}[theorem]{Corollary}
\newtheorem{proposition}[theorem]{Proposition}
\theoremstyle{definition}

\newtheorem{problem}[theorem]{Problem}
\newtheorem{example}[theorem]{Example}
\newtheorem{algorithm}[theorem]{Algorithm}
\theoremstyle{remark}
\DeclareMathOperator{\rk}{rank}
\DeclareMathOperator{\cl}{cl}

\newcommand{\OPT}{\mathrm{OPT}}

\title{Algorithms for Robust Dedicated Actuator Placement}
\author{Qichao Wang\\
\small Hebei University of Technology\\
\small Tianjin, China
\and
Daoyu Dong\\
\small University of Electronic Science\\
\small and Technology of China, Chengdu, China}
\hypersetup{pdftitle={Algorithms for Robust Dedicated Actuator Placement},pdfauthor={Qichao Wang and Daoyu Dong}}
\date{}
\begin{document}
\maketitle
\begin{abstract}
We study the minimum-cost placement of dedicated actuators in a
structured linear system that must remain structurally controllable
after any loss of at most $k$ vulnerable actuators. The state pattern
induces a dual matching matroid of rank $d$, the state matching
deficiency. Robustness separates into resilient spanning in this
matroid and independent accessibility requirements on the source
strongly connected components. For a single failure and unit costs,
a flow relaxation and matching of shared actuator repairs yield a
$3/2$-approximation in $O(n^3)$ graph operations on an $n$-state
digraph. A lower bound based on fundamental cocircuit unions and the
Tutte--Berge formula establishes the ratio. For arbitrary failure
budgets and nonnegative costs, when all actuators are vulnerable,
we give a deterministic exact algorithm in
$2^{O((d+k)^4)}\operatorname{poly}(n,\ell)$ time, where $\ell$
is the total weight encoding length. Its core preserves both
resilient spanning and the minimum cost of completing the source
quotas. For mixed safe and vulnerable locations we give a
one-sided randomized exact optimization algorithm in
$2^{O(kd(d+k)\log(d+1))}\operatorname{poly}(n,\ell,\log(1/\delta))$
time for $d,k\ge1$: every returned placement is feasible, and an
optimum is returned with probability at least $1-\delta$ on feasible
instances. The parameter bounds do not depend on the number of
source components.
\end{abstract}


\section{Introduction}
The placement of actuators determines where external inputs can
influence a dynamical network. When the pattern of state couplings
is known but their numerical values are uncertain, structural
controllability provides a criterion for generic controllability
and a basis for optimizing input locations~\cite{Lin,PKA,PKAcost}.
An actuator configuration designed for the nominal system may lose
controllability after a single input becomes unavailable. We consider
the design of a minimum-cost set of dedicated inputs that continues
to satisfy structural controllability after any permitted set of
actuator failures.

Two requirements govern this design. Each source component of the
state graph needs a surviving input, while the state and input
connections together must admit a matching covering every state.
The first requirement is local to each source component; the second
is a global rank condition. We express the latter as spanning in the
dual of the state matching matroid. Its rank $d$ is precisely the
number of unmatched states in a maximum state matching. The resulting
formulation (Proposition~\ref{prop:control}) combines classical
structural controllability with fault-tolerant matroid optimization.

Nominal minimum dedicated-input placement is polynomial-time
solvable~\cite{PKA}. Its robust counterpart is NP-hard even for a
single failure~\cite{PPR}; hence the failure budget $k$ alone cannot
give fixed-parameter tractability unless $\mathrm P=\mathrm{NP}$.
On the other hand, every robust placement contains a resilient
spanning witness of at most $(k+1)d$ elements. Enumerating all such
witnesses and completing their accessibility requirements gives an
$n^{O((k+1)d)}$ algorithm for positive $d$. The question is whether
one can separate the cost of handling the rank deficit and failures
from the size of the network, even when there are many source
components.

We give three algorithms, summarized in Table~\ref{tab:results}.
Their common basis is the matching--accessibility decomposition.

\begin{enumerate}[leftmargin=*]
\item For one failure and unit costs, we solve a flow relaxation
that already protects accessibility. We then pair critical inputs
that admit a shared backup. The main technical step is a global
lower bound on every resilient spanning set, obtained by applying
the Tutte--Berge formula to a graph of common repairs.
This gives a cubic-time $3/2$-approximation
(Theorem~\ref{thm:three-halves}).

\item For arbitrary failure budgets and nonnegative costs, with all
locations vulnerable, a spanning witness must preserve the cost of
completing source quotas. We introduce adjusted marginal prices
and two nested saturation bounds for each matroid flat, one for
individual quota groups and one for their union. The resulting
bounded core preserves an optimum independently of the number of
groups (Theorem~\ref{thm:quota-fpt}).

\item For mixed safe and vulnerable locations, accessibility becomes
a choice between a safe actuator and a sufficient number of
vulnerable ones. We use a linear representation of the dual
matching matroid and combine coefficient-space compression
with a dynamic program that processes source components one at a
time. Sampling the representation yields a one-sided randomized
algorithm for arbitrary directed state graphs
(Theorem~\ref{thm:random-control}).
\end{enumerate}

\begin{table}[htbp]
\centering\small
\caption{Algorithms for arbitrary directed state graphs. Here
$p=d+k$, $\ell$ is the total cost encoding length, and $\delta$ is the allowed
probability of a suboptimal or unsuccessful randomized trial.}
\label{tab:results}
\begin{tabular}{@{}p{0.29\linewidth}p{0.27\linewidth}p{0.35\linewidth}@{}}
\toprule
Model & Guarantee & Time \\
\midrule
Unit cost, $k=1$ & $3/2$ approximation & $O(n^3)$ graph operations \\
Weighted, all vulnerable & Deterministic optimum & $2^{O(p^4)}\operatorname{poly}(n,\ell)$ \\
Weighted, mixed failures & Optimum with probability $\ge 1-\delta$ &
$2^{O(kdp\log(d+1))}$\newline$\cdot\operatorname{poly}(n,\ell,\log(1/\delta))$ \\
\bottomrule
\end{tabular}
\end{table}

\paragraph{Related work.}
Lin~\cite{Lin} established the graph characterization of structural
controllability. Pequito, Kar, and Aguiar~\cite{PKA,PKAcost} developed
minimum dedicated-input placement and weighted input/output
configuration design. Pequito et al.~\cite{PRKAP} study robust
controllability for numerical systems, and Guo et al.~\cite{GuoBackup}
optimize backup actuators for a prescribed primary placement.
Po\c{c}as, Pequito, and Ramos~\cite{PPR} establish NP-hardness of
robust minimal structural controllability and give cubic-time
approximation ratios $5/2$ for general graphs and $2$ for strongly
connected graphs in the one-failure dedicated model. Our
Theorem~\ref{thm:three-halves} improves both bounds for the same
distinct-state, unit-cost model. Related repair strategies occur
in node-robust and edge-robust assignment~\cite{ABMnodes,ABMedges}.

On the matroid side, Bentert et al.~\cite{BFGL} give fixed-parameter
algorithms for unweighted fault-tolerant spanning. Kundu, Nandi,
and Saurabh~\cite{KNS} address the weighted problem with an
independence oracle, including mixed failure statuses. Their
quartic-exponent algorithm uses flat saturation, and their linear
matroid algorithm uses coefficient-space compression. We build on
these methods to handle the accessibility requirements of
general state graphs. The deterministic extension preserves the
cost of quota completion, while the represented-matroid extension
incorporates a group-survival state in the optimization and
transfers the result to directed graphs.

Section~\ref{sec:model} establishes the control formulation and
illustrates the two constraints. Section~\ref{sec:approx} treats
single failures; Sections~\ref{sec:quota-fpt} and
\ref{sec:mixed} treat the weighted problems. Section~\ref{sec:conclusion}
discusses the model, the parameter regime, and reproducibility.

\section{System model and graph characterization}\label{sec:model}
\subsection{Linear dynamics and actuator placement}
Consider a continuous-time linear time-invariant system
\begin{equation}\label{eq:dynamics}
 \dot{x}(t)=Ax(t)+B_Pu(t),\qquad x(t)\in\mathbb R^n.
\end{equation}
The state matrix has a prescribed structural pattern
$\bar A\in\{0,\star\}^{n\times n}$: a zero entry is fixed to
zero and each $\star$ entry represents a separate free real
parameter. We assume $n\ge1$. A numerical realization of a
pattern is a real matrix obtained by assigning values to these
parameters. Write $E=\{1,\ldots,n\}$ for the state indices and
$e_i$ for the $i$th coordinate vector of $\mathbb R^n$.

A placement $P=\{p_1,\ldots,p_\kappa\}\subseteq E$,
where $\kappa=|P|$, installs one dedicated scalar input at each selected state. Its
input matrix is
\begin{equation}\label{eq:dedicated}
 B_P=[b_{p_1}e_{p_1}\ \cdots\ b_{p_\kappa}e_{p_\kappa}],
 \qquad u(t)\in\mathbb R^\kappa,
\end{equation}
where the coefficients $b_{p_j}$ are free parameters independent
of the entries of $A$. All states are candidate locations,
and at most one actuator may be installed at each state.
The notation $\bar B_P$ denotes the corresponding structural
pattern.

A numerical pair $(A,B_P)$ is controllable if its Kalman
controllability matrix has full row rank:
\[
 \operatorname{rank}[B_P,\ AB_P,\ldots,A^{n-1}B_P]=n.
\]
The structural pair $(\bar A,\bar B_P)$ is structurally
controllable if there exists a controllable numerical
realization. For independent free parameters this is equivalent
to controllability for almost every realization~\cite{Lin,PKA},
because the full-rank condition is expressed by nonzero polynomial
minors of the controllability matrix.

An actuator failure at state $i\in P$ removes its input column
from~\eqref{eq:dedicated}. Thus for a set $F\subseteq P$ of
failed actuators, the surviving pair is
$(\bar A,\bar B_{P\setminus F})$; the state matrix and its
pattern remain unchanged. For a specified vulnerable set $V\subseteq E$, we call $P$
$k$-robust if the surviving pair is structurally controllable for
every $F\subseteq P\cap V$ with $|F|\le k$. Since the family of permitted failures is
finite, structural controllability of each surviving pattern
also implies simultaneous controllability of all survivors
for almost every common assignment of the free coefficients.
The placement is chosen before $F$ is known. When costs
$w_i\ge0$ are specified, the objective is
$\min_{P}\sum_{i\in P}w_i$ subject to $k$-robustness.
For unit costs this is minimum cardinality. In the mixed
failure model a set $V\subseteq E$ specifies the vulnerable locations,
so permitted failures satisfy $F\subseteq P\cap V$ and $|F|\le k$.
The all-vulnerable model corresponds to $V=E$.

\subsection{From state couplings to matching and accessibility}
The state graph $G=(E,\mathcal A)$ has an arc $j\to i$
exactly when $\bar A_{ij}=\star$; self-loops are allowed.
For a placement $P$, the input-state digraph additionally
has one dedicated input vertex $u_p$ and one arc $u_p\to p$
for each $p\in P$. The state bipartite graph has a left and a
right copy of $E$, with an edge $j_L i_R$ for each state arc
$j\to i$. The input-augmented bipartite graph adds the input
vertices on the left and the edges $u_p p_R$. The classical graph criterion is the starting point of our reduction.

\begin{lemma}[Structural controllability criterion~\cite{Lin,PKA}]
\label{lem:classical}
A structured pair is structurally controllable if and only if
every state is reachable from an input in its input-state digraph,
and its input-augmented bipartite graph has a matching covering
all right-state vertices.
\end{lemma}

The first condition transmits input influence through the graph;
the second excludes structural dilations.

\subsection{The dual matching matroid}

Write $S_1,\ldots,S_s$ for the source strongly connected components (SCCs),
those having no entering state arc from another SCC. Let $\mathcal T$ be the transversal matroid on the right
copies: its independent sets are the right-state sets matchable by
bipartite edges. Put $M=\mathcal T^*$ and
$d=n-\rk_{\mathcal T}(E)$, the maximum-matching deficiency.
We identify each state, its right copy, and the corresponding element of
$M$ when no ambiguity can arise. We write $\rk_M$ when the matroid
must be specified, and $\rk$ when it is clear from context.
In general matroid statements the rank is $r$; for this control
construction, $r=d$.
For $X\subseteq E$, duality gives
\begin{equation}\label{eq:dual-rank}
 \rk_M(X)=|X|+\rk_{\mathcal T}(E\setminus X)-\rk_{\mathcal T}(E).
\end{equation}
In particular, $X$ spans $M$ if and only if $E\setminus X$ is independent
in $\mathcal T$. A rank oracle for $M$ is computable by bipartite matching.

We use \emph{resilient spanning} for the matroid analogue of
$k$-robustness: after each permitted failure the remaining elements
must still span the matroid.

The two graph conditions of Section~\ref{sec:model} now give the
following placement criterion.
\begin{proposition}[Matching--accessibility decomposition]\label{prop:control}
A placement $P$ is $k$-robust for vulnerable set $V$ if and only if
\begin{align}
 \rk_M(P\setminus F)&=d
 &&(F\subseteq P\cap V,\ |F|\le k),\label{eq:control}\\
 P\cap(S_i\setminus V)&\ne\varnothing
 \quad\text{or}\quad |P\cap S_i\cap V|\ge k+1
 &&(1\le i\le s).\label{eq:access}
\end{align}
For $V=E$, \eqref{eq:access} is simply $|P\cap S_i|\ge k+1$.
\end{proposition}
\begin{proof}
Fix the surviving placement $X=P\setminus F$.
A matching of $E\setminus X$ using state edges can be extended by
matching each state in $X$ to its own input. Conversely, a covering
input-augmented matching must match $E\setminus X$ using state edges,
because these states have no incident input edges. Equation~\eqref{eq:dual-rank}
therefore makes the covering-matching condition equivalent to spanning
in $M$.

Every component in the condensation digraph is reachable from a
source component, and no input outside a source component can reach
that component. Thus accessibility is equivalent to $X$ meeting each
source SCC. If a source contains a selected safe input, or at least
$k+1$ selected vulnerable inputs, it retains an input under every
permitted failure. Otherwise failing its at most $k$ selected inputs
violates accessibility. Applying Lemma~\ref{lem:classical} proves both
directions.
\end{proof}

\begin{problem}[Robust dedicated-actuator placement]\label{prob:control}
Given $\bar A$, costs $w_i\ge0$, a vulnerable set $V$ (default $E$),
and failure budget $k$, minimize $\sum_{i\in P}w_i$ subject to
\eqref{eq:control} and~\eqref{eq:access}, or certify infeasibility.
\end{problem}

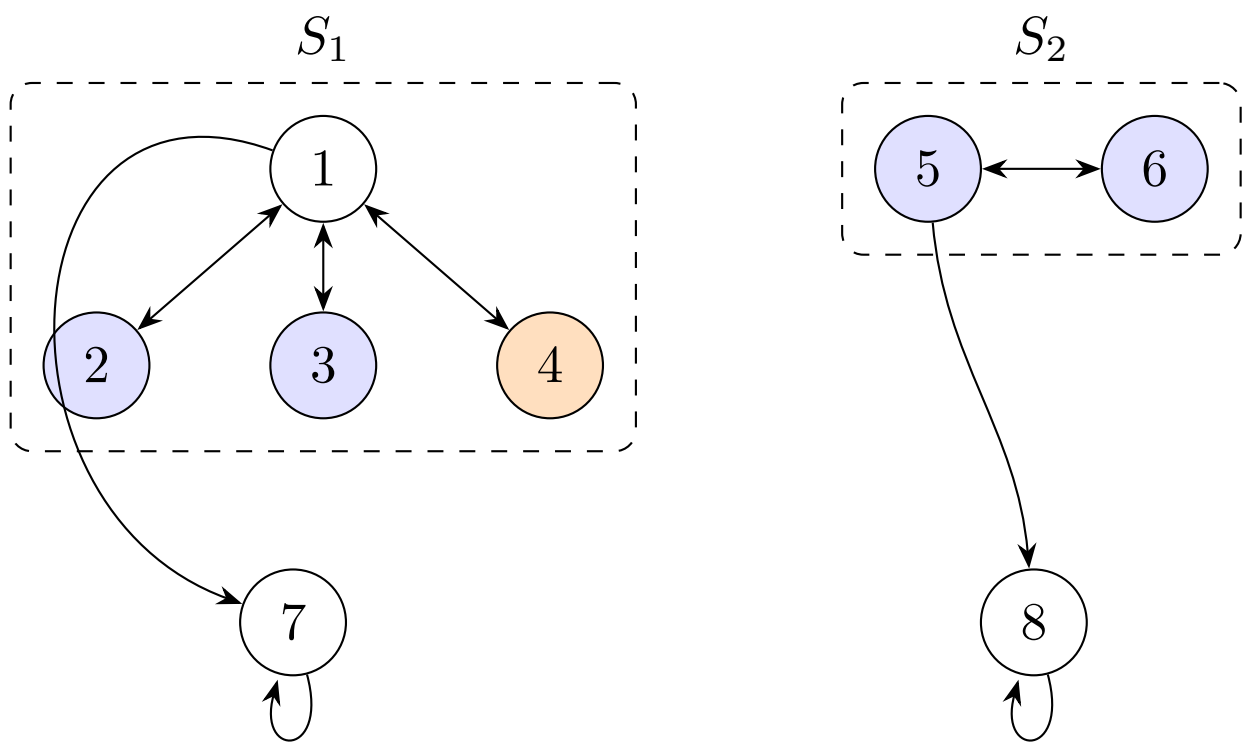
\begin{figure}[htbp]
\centering
\begin{tikzpicture}[>=Stealth, v/.style={circle,draw,minimum size=7mm,inner sep=0pt},font=\small]
\node[v] (a) at (0,0) {1};
\node[v,fill=blue!12] (b) at (-1.5,-1.3) {2};
\node[v,fill=blue!12] (c) at (0,-1.3) {3};
\node[v,fill=orange!25] (d) at (1.5,-1.3) {4};
\node[v,fill=blue!12] (e) at (4,0) {5};
\node[v,fill=blue!12] (f) at (5.5,0) {6};
\node[v] (g) at (-.2,-3) {7};
\node[v] (h) at (4.7,-3) {8};
\draw[<->] (a)--(b);\draw[<->] (a)--(c);\draw[<->] (a)--(d);
\draw[<->] (e)--(f);
\draw[->] (a) to[out=160,in=160,looseness=1.6] (g);
\draw[->] (e) to[out=-85,in=95] (h);
\draw[->] (g) edge[loop below] (g);\draw[->] (h) edge[loop below] (h);
\node[draw,dashed,rounded corners,fit=(a)(b)(c)(d),inner sep=6pt,label=above:{$S_1$}] {};
\node[draw,dashed,rounded corners,fit=(e)(f),inner sep=6pt,label=above:{$S_2$}] {};
\end{tikzpicture}
\caption{Running example. Shaded blue states form a minimum
source-quota relaxation for one failure; the orange state is a
shared backup. Dashed boxes mark the source SCCs.}
\label{fig:running}
\end{figure}

\paragraph{A running example.}
Figure~\ref{fig:running} shows a network with eight states and two
source SCCs, $\{1,2,3,4\}$ and $\{5,6\}$. States 7 and 8 have
self-loops and incoming edges from states 1 and 5, respectively.
The first component is a bidirected star. The other source component
is a bidirected pair. The state matching has size six: at most one
leaf of the star can be matched by a state edge, while the other five
nonleaf states can all be matched. Hence $d=2$, and the dual matroid
has rank two on $\{2,3,4\}$, with every pair of these leaves a basis
and every other element a loop. This follows because the complement
of any two leaves admits a state matching, and a maximum state
matching must leave two leaves unmatched.

Inputs at states $\{1,5\}$ reach every state but fail the matching
condition. Inputs at $\{2,3\}$ supply the matching deficit but
cannot reach the second source component. For $k=1$, the optimal
placement contains all three leaves and both states of the second
source component. We will recover this five-input solution by
both the approximation and the exact algorithms.

\paragraph{Interpretation of the model and parameter.}
The distinct-location assumption follows the dedicated model
of~\cite{PPR}. It is appropriate when the design variable is an
actuation location, or when multiple devices at one location are
treated as one fault domain. It excludes independent redundant
devices at an identical state; allowing such devices changes the
feasible family. If every state has a self-loop, the state graph
has a perfect matching and $d=0$. The problem then separates into
independent source-component decisions. Conversely, the example
in Figure~\ref{fig:running} can be extended by arbitrarily many
downstream states with self-loops while retaining $d=2$ and a
nontrivial redundancy requirement. This gives a family of
large systems with bounded positive matching deficiency.

\paragraph{Matroid notation.}
For a finite matroid, $\rk$ denotes rank and
$\cl(X)=\{e:\rk(X\cup\{e\})=\rk(X)\}$ denotes closure.
A flat equals its closure; a hyperplane is a flat of rank one
below the ambient rank. A loop has rank zero; a coloop belongs
to every basis. A circuit is a minimally dependent set.
For a basis $B$ and $e\notin B$, its fundamental circuit is the
unique circuit in $B\cup\{e\}$. The fundamental cocircuit at
$b\in B$ is $E\setminus\cl(B\setminus\{b\})$~\cite{Oxley}.
We use $U_{r,m}$ for the uniform matroid of rank $r$ on $m$ elements.
The term \emph{placement} refers to the control input locations,
and \emph{resilient witness} to a matroid subset that spans after
every permitted failure.

\section{A cubic-time 3/2-approximation}\label{sec:approx}
Throughout this section $k=1$, all actuators are vulnerable, and
every actuator has unit cost.
By Proposition~\ref{prop:control}, every source component needs two
inputs and the selected states must span the matching matroid even
after deleting one of them. We enforce the accessibility requirement
at the first stage; the second stage adds backup inputs
to repair the possible losses of spanning. The cost of this
augmentation is bounded against the globally optimal robust
placement. This separates the easy local constraint from the
global rank condition. A critical input is one whose loss destroys
spanning; its cocircuit lists possible replacements. Two such inputs
can share a backup even though they cannot fail simultaneously.
The matching of pairs yields the augmentation cost. To bound it
against the unknown optimum, we count how many elements a robust
solution must take from each disjoint region of cocircuits.

\subsection{An exact source-quota relaxation}
\begin{proposition}\label{prop:one-failure-feasibility}
In the one-failure model, a robust placement exists if and only
if every source SCC contains at least two states. In this case the dual
matching matroid $M$ has no coloops.
\end{proposition}
\begin{proof}
A singleton source SCC cannot retain an actuator after one failure,
since at most one actuator may be installed at each state.
Conversely, if all source SCCs have at least two vertices, every state has
an incoming state arc: this holds inside each nonsingleton SCC, while
a singleton nonsource SCC has an entering arc. Thus $\mathcal T$ has no
loops and its dual $M$ has no coloops. The full placement $E$ spans
after any single deletion and satisfies all source quotas.
\end{proof}

Minimize $|P|$ subject to $P$ spanning $M$ and $|P\cap S_i|\ge2$ for
all $i$. Let $L$ be the optimum of this relaxation and $\OPT$ the minimum
cardinality of a robust placement. Every robust placement is feasible
for the relaxation, so $L\le\OPT$.

By Proposition~\ref{prop:one-failure-feasibility}, we may reject
instances with a singleton source SCC and assume $|S_i|\ge2$ for every $i$. The relaxation then
reduces to an integral maximum-flow problem.
The complement $I=E\setminus P$
must be matchable using state edges and must satisfy
$|I\cap S_i|\le |S_i|-2$. The network has a super-source, two copies of the state set $E$, one group
node per source SCC, and a super-sink. Edges from the super-source to
left-state vertices and bipartite state edges have unit capacity.
Each right-state vertex has a unit-capacity outgoing edge to its group
node if it belongs to a source SCC, and to the super-sink otherwise.
The edge from group node $i$ to the super-sink has capacity
$|S_i|-2$. An integral maximum flow gives a maximum-cardinality admissible
$I$, and therefore a minimum-cardinality $P$.

\subsection{Cocircuit unions and a global lower bound}
The next two lemmas concern an arbitrary rank-$d$ matroid $M$ with a
spanning set $P$. Let $D$ be the coloops of $M|P$, calling them
critical elements and the elements of $P\setminus D$ noncritical.
Every critical element belongs to every basis contained in $P$.
For $b\in D$ define
\[
 C_b=E\setminus\cl(P\setminus\{b\}).
\]
The set $C_b$ is the fundamental cocircuit at $b$ with respect to
any basis of $P$. It meets $P$ exactly in $\{b\}$, and an external
element of $C_b$ restores spanning after deleting $b$.

\begin{lemma}\label{lem:union-rank}
For every $R\subseteq D$,
\[
 \rk\Bigl(E\setminus\bigcup_{b\in R}C_b\Bigr)=d-|R|.
\]
\end{lemma}
\begin{proof}
Choose a basis $B\subseteq P$. Every $b\in D$ belongs to $B$.
Since $B\setminus\{b\}\subseteq P\setminus\{b\}$ and both sets have
rank $d-1$, their closures coincide. For nonempty $R$ we claim
\[
 \bigcap_{b\in R}\cl(B\setminus\{b\})=\cl(B\setminus R).
\]
For $e\notin B$, membership in $\cl(B\setminus\{b\})$ means that its
fundamental circuit omits $b$. Membership for every $b\in R$ therefore
means that this circuit is contained in $(B\setminus R)\cup\{e\}$,
which is precisely membership in $\cl(B\setminus R)$. Basis elements
satisfy the equality directly, and taking ranks proves the lemma. For $R=\varnothing$, the
statement reduces to $\rk(E)=d$.
\end{proof}

Form the \emph{common-repair graph} $K$, a simple undirected
(possibly nonbipartite) graph on $D$, with an edge $bc$ when some
$e\in E\setminus P$ belongs to both $C_b$ and $C_c$.
Let $\nu$ denote its maximum matching cardinality. A backup may
repair more than two critical inputs; the graph construction uses
pairwise sharing and the matching analysis guarantees the cost of
this particular augmentation.

In Figure~\ref{fig:running}, the relaxation can return
$P=\{2,3,5,6\}$. Then $D=\{2,3\}$, $C_2=\{2,4\}$,
$C_3=\{3,4\}$, and $K$ is a single edge witnessed by state 4.
The algorithm adds that state and returns the five-input optimum.
The following bound explains why pairing critical inputs suffices
even when the initial relaxation is far from a robust solution.

\Needspace{6\baselineskip}
\begin{lemma}\label{lem:packing}
Every set $X$ spanning $M$ after any one deletion satisfies
$|X|\ge2(|D|-\nu)$.
\end{lemma}
\begin{proof}
Fix $U\subseteq D$ and let $R_1,\ldots,R_t$ be the nonempty connected
components of $K-U$. Put $Q_i=\bigcup_{b\in R_i}C_b$. These sets are
pairwise disjoint. An element of $P$ belongs to at most one $C_b$. A common
external element of $Q_i,Q_j$ would witness an edge between their components.
Lemma~\ref{lem:union-rank} gives $\rk(E\setminus Q_i)=d-|R_i|$.
Since adding an element raises rank by at most one, $X$ must contain at
least $|R_i|$ elements of $Q_i$ to span. Equality is impossible: since $R_i$ is nonempty, equality gives at
least one element in $X\cap Q_i$. Deleting one of these elements would leave rank at most
$(d-|R_i|)+(|R_i|-1)<d$. Thus
\Needspace{5\baselineskip}
\[
 |X|\ge\sum_i(|R_i|+1)=|D|-|U|+t
 \ge |D|-|U|+\operatorname{odd}(K-U),
\]
where $\operatorname{odd}$ counts odd-order components. The classical
Tutte--Berge formula~\cite{LP} states
\[
 2\nu=|D|-\max_{U\subseteq D}
 \bigl(\operatorname{odd}(K-U)-|U|\bigr).
\]
Taking the maximum in the preceding inequality proves the result.
\end{proof}

\subsection{Algorithm and guarantee}
\begin{algorithm}[Common-repair augmentation]\label{alg:repair}
\begin{enumerate}[label=\arabic*.,leftmargin=*]
\item Reject if a singleton source SCC exists. Otherwise, solve the
source-quota flow relaxation for an optimal set $P$.
\item Find the coloops $D$ of $M|P$, construct $K$, and retain one
external common-repair witness for each edge of $K$.
\item Find a maximum matching in $K$. Select the stored witness for each
matched pair and any external element of $C_b$ for each unmatched
$b\in D$. Return $P$ together with these witnesses, removing duplicates.
\end{enumerate}
\end{algorithm}
\begin{theorem}\label{thm:three-halves}
For any nonempty directed state graph in the stated dedicated-actuator
model, a robust placement of size at most $\tfrac32\OPT$ can be found, or
infeasibility certified, in $O(n^3)$ elementary graph operations
and $O(n^2)$ space.
The guarantee is for unit costs and one actuator failure.
\end{theorem}
\begin{proof}
By Proposition~\ref{prop:one-failure-feasibility}, Algorithm~\ref{alg:repair}
rejects exactly the infeasible instances. For a feasible instance, let
$P$ be an optimal solution of the flow relaxation and let $R_{\rm rep}$
be the selected witness set, with duplicates removed. Every unmatched
$b\in D$ has an external witness because $M$ has no coloops while
$P\setminus\{b\}$ has rank $d-1$. At most
$\nu+(|D|-2\nu)=|D|-\nu$ witnesses are selected.

The set $P\cup R_{\rm rep}$ is robust. If a witness fails, $P$ still spans. If an
original noncritical element fails, the surviving part of $P$ spans. If
$b\in D$ fails, its witness repairs the deficiency. The source quotas
persist under additions. By Lemma~\ref{lem:packing},
\[
 |P\cup R_{\rm rep}|\le L+|D|-\nu\le\OPT+\frac{\OPT}{2}.
\]
For the complexity, retain the state matching $J$ obtained from the quota
flow. It covers $I=E\setminus P$. For each $b\in P$, explore alternating
paths from unmatched right vertex $b$: traverse state edges to the left,
then matched edges back to the right. An alternating path reaching a free left vertex is
an augmenting path witnessing that $I\cup\{b\}$ is matchable, so
$b$ is noncritical. For a critical $b$, the reachable matched right
vertices are exactly its external repairs. A path to right vertex $e$
can be flipped to match $(I\setminus\{e\})\cup\{b\}$. Conversely, the
symmetric difference with a matching of that replacement set contains
an alternating path starting at $b$. Its endpoint is either the displaced
right vertex $e$ or a left vertex free under $J$. The latter endpoint
would give an augmentation and contradict criticality.
Hence these searches recover the complete repair supports.

There are $O(n)$ vertices and $O(n^2)$ edges in the flow network. At most
$n$ unit augmentations, each found in $O(n^2)$ operations, suffice. At most
$n$ alternating searches cost $O(n^3)$ in total. For each external state,
connect all pairs of critical states it repairs, storing a witness for each
edge. This clique construction costs $O(n^3)$. A maximum general-graph
matching is computable in $O(n^3)$ operations by a blossom algorithm~\cite{Gabow,LP}. All structures require $O(n^2)$ space.
\end{proof}

\begin{example}[Three-state bidirected star]\label{ex:star}
Consider the three-state bidirected star with center $a$
and leaves $b,c$.
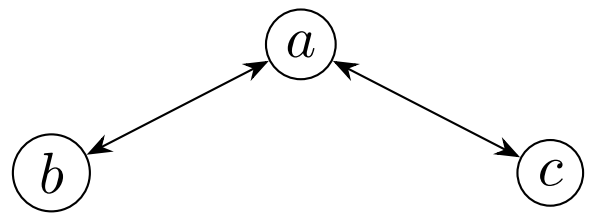
\begin{center}
\begin{tikzpicture}[>=Stealth,baseline=(current bounding box.center)]
\node[draw,circle,inner sep=2pt] (a) at (0,0) {$a$};
\node[draw,circle,inner sep=2pt] (b) at (-1.65,-0.85) {$b$};
\node[draw,circle,inner sep=2pt] (c) at (1.65,-0.85) {$c$};
\draw[<->] (a) -- (b);
\draw[<->] (a) -- (c);
\end{tikzpicture}
\end{center}
The relaxation may choose $P=\{a,c\}$, whose only critical state is
$c$ and whose only external repair is $b$. Algorithm~\ref{alg:repair}
returns all three states, whereas $\{b,c\}$ is robust.
Thus the $3/2$ guarantee is attained under worst-case tie-breaking
for Algorithm~\ref{alg:repair} as stated. A subsequent redundancy
deletion would remove $a$ in this example; the tightness concerns
the algorithm without that postprocessing.
\end{example}

\section{Exact optimization with source quotas}\label{sec:quota-fpt}
The approximation algorithm protects accessibility before repairing
the matching condition. For exact optimization we reverse this
order: find a small resilient spanning witness, then complete the
source quotas at minimum cost. In Figure~\ref{fig:running}, the
witness $C=\{2,3,4\}$ already spans after one failure. Its completion
requires the two inputs at states 5 and 6, giving five inputs.
An arbitrary spanning core cannot be optimized in isolation:
in $U_{1,4}$ with $k=1$ and a single quota-two group $\{c,d\}$,
the witness $\{a,b\}$ needs two additional elements, whereas
$\{c,d\}$ needs none. We therefore preserve the completion cost.

Throughout this section let $M$ be a rank-$r$ matroid on $E$ given
by an independence oracle, $n=|E|$, $V\subseteq E$ the vulnerable
elements, and $k\ge0$. A witness $C$ is resilient if
$\rk(C\setminus F)=r$ for all $F\subseteq C\cap V$ with $|F|\le k$.
For $r\ge1$ write
\[
 b=(k+1)r,\qquad H=\binom{r+k-1}{k}.
\]
The following lemmas include the rank-one and zero-failure cases.

\subsection{Small resilient witnesses}
The layered bound below follows the construction of Bentert
et al.~\cite[Proposition~6]{BFGL}.
\begin{lemma}\label{lem:small}
Every resilient set $B$ contains a resilient subset $B_0$ of cardinality at most $b$.
\end{lemma}
\begin{proof}
Successively choose a maximal independent subset $I_j$ of $B\setminus\bigcup_{i<j}I_i$, for $j=0,\ldots,k$, allowing empty layers, and put
$B_0=\bigcup_{j=0}^{k}I_j$. Since $|I_j|\leq r$, we have $|B_0|\leq b$.

Fix $F\subseteq B_0\cap V$ with $|F|\leq k$. The layers are disjoint, so some $I_j$ avoids $F$. Every element of $B\setminus B_0$ was available when $I_j$ was chosen and belongs to $\cl(I_j)$. Every element of $B_0\setminus F$ already lies in that surviving set. Therefore
\[
 B\setminus F\subseteq \cl(B_0\setminus F).
\]
Since $F$ is also an allowed failure of $B$, its left-hand side spans $M$. Thus $B_0$ is resilient. Removing matroid loops from $B_0$ preserves resilience.
\end{proof}

\subsection{Deficient hyperplanes and resilience-preserving exchanges}
The set-pairs inequality of Bollob\'as~\cite{Bollobas}
sharpens the hyperplane count in Bentert et al.~\cite[Lemma~8]{BFGL}.

\begin{lemma}\label{lem:hyperplanes}
For $r\ge1$, $k\ge0$, and any $Y\subseteq E$, at most $\binom{r+k-1}{k}$ distinct hyperplanes can have the form $\cl(Y\setminus F)$ with $F\subseteq Y$ and $|F|\leq k$. This count is sharp.
\end{lemma}
\begin{proof}
If $r=1$, the closure of the empty set is the unique hyperplane.
If $k=0$, only $F=\varnothing$ is allowed. In either case the
bound equals one; it is attained in the uniform-matroid example
below. In general write
$\Pi_1,\ldots,\Pi_m$ for these distinct hyperplanes. Choose a basis $A_i\subseteq Y\cap \Pi_i$ of size $r-1$, and put $\Delta_i=Y\setminus \Pi_i$. Because $\Pi_i=\cl(Y\setminus F)$ for some permitted $F$, we have $\Delta_i\subseteq F$, so $|\Delta_i|\leq k$. Also $A_i\cap \Delta_i=\varnothing$. If $i\ne j$, then $A_i\cap \Delta_j\ne\varnothing$: otherwise $A_i\subseteq \Pi_j$ and hence $\Pi_i=\cl(A_i)\subseteq \Pi_j$, forcing equality of the two hyperplanes.

In a uniformly random permutation of $Y$, let $\mathcal E_i$ be the event that every element of $A_i$ precedes every element of $\Delta_i$. Its probability is $\binom{r-1+|\Delta_i|}{r-1}^{-1}$. These events
are pairwise disjoint. Indeed, for $i\ne j$ choose $a\in A_i\cap \Delta_j$ and $b\in A_j\cap \Delta_i$. The two events require $a$ before $b$ and $b$ before $a$, respectively.
Therefore
\Needspace{5\baselineskip}
\[
 \sum_i\binom{r-1+|\Delta_i|}{r-1}^{-1}\leq1.
\]
Each summand is at least $\binom{r+k-1}{r-1}^{-1}$, proving the bound. For sharpness in all cases take $M=U_{r,r+k}$, $B=E$, all elements vulnerable, and $Y=B\setminus\{x\}$. The set $B$ is resilient, and each $k$-subset of $Y$ leaves a distinct $(r-1)$-element hyperplane. This gives exactly $\binom{r+k-1}{k}$ distinct hyperplanes of the required form.
\end{proof}

\begin{lemma}[Exchange outside deficient hyperplanes]\label{lem:exchange}
Let $C$ be resilient and $x\in C$, and put $Y=C\setminus\{x\}$.
For each permitted failure $F\subseteq Y\cap V$, $|F|\le k$,
with $\rk(Y\setminus F)<r$, form $\Pi_F=\cl(Y\setminus F)$.
These are at most $H=\binom{r+k-1}{k}$ distinct hyperplanes,
all avoiding $x$. Any $y\notin C$ of the same failure status as $x$
that avoids all these hyperplanes makes $Y\cup\{y\}$ resilient.
\end{lemma}
\begin{proof}
Because $C$ is resilient, adding $x$ restores full rank after any
such $F$. Each deficient survivor has rank $r-1$ and its closure
avoids $x$. Lemma~\ref{lem:hyperplanes} bounds their number.
If $y$ survives a permitted failure of the replacement set,
it restores any deficient survivor of $Y$. If $y$ fails, replace
its failure by that of $x$ in $C$; the resulting survivor is
identical and the failure is permitted because $x$ and $y$ have
the same status.
\end{proof}

\subsection{Minimum-cost completion}
Let $S_1,\ldots,S_s$ be disjoint quota groups, with lower
bounds $\beta_i\ge1$, and let $S_0=E\setminus\bigcup_iS_i$.
The cost of a set is $w(P)=\sum_{e\in P}w(e)$, with nonnegative
binary-encoded weights. We reject a group with fewer than
$\beta_i$ elements. For any forced set $C$, write $g(C)$ for
the minimum cost of a superset of $C$ meeting all quotas.
It is obtained by buying $C$ and adding the cheapest
$\max(0,\beta_i-|C\cap S_i|)$ unused elements of each group.
Loops remain available for completion, although they are not
needed in a spanning witness.

By Lemma~\ref{lem:small} the optimum is
\begin{equation}\label{eq:completion}
 \min\{g(C): C\text{ resilient},\ |C|\le b,\ C\text{ loopless}\}.
\end{equation}
Indeed a feasible purchase contains such a witness; conversely,
every quota-completing superset of a resilient witness is resilient.

In the unit-cost case with $\beta_i=k+1=\beta$, the completion
cardinality is
\[
 g(C)=\beta s+|C\cap S_0|+
      \sum_{i=1}^s\max(0,|C\cap S_i|-\beta).
\]
For weighted inputs the marginal price depends on which elements
are already forced. The next lemma supplies a static lower bound
on that marginal which is exact for an untouched group.

For $i\ge1$ let $a_i$ be the $\beta_i$th cheapest weight in $S_i$,
and define
\[
 v(x)=
 \begin{cases}
 \max(0,w(x)-a_i),& x\in S_i,\ i\ge1,\\
 w(x),&x\in S_0.
 \end{cases}
\]
\begin{lemma}[Completion prices]\label{lem:prices}
For $x\notin Y$,
$g(Y\cup\{x\})-g(Y)\ge v(x)$, with equality if $x\in S_0$ or
$Y$ avoids the group containing $x$.
If $x,y\notin Y$ belong to the same group (including $S_0$)
and $w(y)\le w(x)$, then
$g(Y\cup\{y\})\le g(Y\cup\{x\})$.
\end{lemma}
\begin{proof}
In a group with an unfilled quota put $t=\beta_i-|Y\cap S_i|$,
and let $c$ be the $t$th cheapest weight in $S_i\setminus Y$.
At least $t$ of the original $\beta_i$ cheapest elements remain,
so $c\le a_i$. Forcing $x$ changes the optimum cost by
$\max(0,w(x)-c)$; if the quota is filled the change is $w(x)$.
For an untouched group $c=a_i$, and for $S_0$ costs are additive.
This proves the marginal bound. For the second claim take a
minimum-cost completion of $Y\cup\{x\}$. If it contains $y$,
it is also a completion of $Y\cup\{y\}$. Otherwise replace $x$
by $y$, retaining all quota counts at no greater cost.
\end{proof}

\subsection{A bounded core preserving completion cost}
Kundu et al.~\cite[Lemma~5]{KNS} use a rank-dependent capacity for
every flat to retain cheap elements of a resilient spanning set.
We extend this rule to two capacities: one bounds the retained
elements of any individual group, and the other bounds their total.
The two types of saturation yield different exchanges.

\begin{theorem}[Weighted disjoint quotas]\label{thm:quota-fpt}
A minimum-weight resilient set satisfying all disjoint cardinality
quotas, or infeasibility, can be found deterministically in
\[
 2^{O(b(\log(b+1)+r\log(H+1)))}\operatorname{poly}(n,\ell)
\]
time for $r,k\ge1$, in particular
$2^{O((r+k)^4)}\operatorname{poly}(n,\ell)$ time, where $\ell$ is the total binary weight
encoding length. The exponential factor is independent of $s$.
For $r=0$ and for $k=0$, the problem is polynomial-time solvable.
\end{theorem}
\begin{proof}
For $r=0$, choose the cheapest quota completions. For $k=0$,
the complement of a feasible purchase is independent in the dual
matroid and obeys the upper bounds $|S_i|-\beta_i$ on its groups.
Maximum-weight matroid intersection with a partition matroid
solves this case~\cite{Lawler,Oxley}.

Assume $r,k\ge1$. Set $L_0=A_0=0$ and, for $h=1,\ldots,r$,
\[
 L_h=HL_{h-1}+b,\qquad A_h=HA_{h-1}+bL_h+b.
\]
Process the safe and vulnerable nonloops in separate scans, in
lexicographic order $(v(x),w(x),x)$, using any fixed order of
element identifiers. Within each scan retain a set $W$ with
\begin{equation}\label{eq:flat-caps}
 |W\cap Q\cap S_i|\le L_{\rk(Q)}\quad(0\le i\le s),
 \qquad |W\cap Q|\le A_{\rk(Q)}
\end{equation}
for every ambient flat $Q$. Reject an incoming $x$ if a flat
containing it is saturated in the total cap or in the cap of its
own group; otherwise retain it. The invariants hold because only
the total and own-group counts can increase. Each scan retains
at most $A_r$ elements.

\paragraph{Exchange.}
Fix a resilient loopless $C$ of size at most $b$, and an element
$x\in C$ rejected by its scan. Put $Y=C\setminus\{x\}$.
Lemma~\ref{lem:exchange} identifies at most $H$ deficient
hyperplanes avoiding $x$. In any saturated witness flat $Q$
of rank $h$ containing $x$, each of these hyperplanes intersects
$Q$ in a flat of rank at most $h-1$. We use retained elements
present at the time of rejection.

If the own-group cap is saturated, there are $L_h$ candidates
in $Q$ from the group of $x$. At most $HL_{h-1}$ belong to
the deficient hyperplanes, so at least $b$ avoid all of them.
At most $b-1$ belong to $Y$. Choose another candidate $y$.
It has the same failure status as $x$, and the ordering within
a group implies $w(y)\le w(x)$. By Lemmas~\ref{lem:exchange}
and~\ref{lem:prices}, the replacement preserves resilience and
does not increase completion cost.

If the total cap is saturated, exclude $S_0$ and every group
touched by $Y$. There are at most $|Y|+1\le b$ such groups,
each with at most $L_h$ candidates. The deficient hyperplanes
cover at most $HA_{h-1}$ further candidates. At least
\[
 A_h-bL_h-HA_{h-1}=b
\]
remain. Each is in a group untouched by $Y$, hence outside $Y$,
and has the same failure status as $x$. For such a candidate $y$,
the scan order gives $v(y)\le v(x)$, and Lemma~\ref{lem:prices} gives
\[
 g(Y\cup\{y\})-g(Y)=v(y)\le v(x)
 \le g(Y\cup\{x\})-g(Y).
\]
Resilience follows from Lemma~\ref{lem:exchange}. In both cases
the replacement lies in the final retained set. Repeating reduces
the number of elements of $C$ outside the union of the two scans
until none remain. Thus the union contains a witness attaining
\eqref{eq:completion}.

\paragraph{Implementation and complexity.}
Every saturated flat is spanned by its relevant retained
elements. Otherwise the closure of those elements is a flat
of lower rank with the same count, contradicting the strictly
smaller cap. Consequently a scan can detect all saturation
witnesses by enumerating independent subsets of the retained
set of size at most $r$ and testing closure membership and
the two counts. The scan costs
$n(2A_r+1)^{O(r+1)}\operatorname{poly}(n,\ell)$ operations
and oracle calls. The recurrences give
\[
 L_r=b\sum_{j=0}^{r-1}H^j,\qquad
 A_r=L_r+b^2\sum_{j=0}^{r-1}(j+1)H^j .
\]
Thus $\log(2A_r+1)=O(\log(b+1)+r\log(H+1))$.
Enumerate all subsets of the union core of size at most $b$,
test the at most $2^b$ permitted failures by rank queries, and
complete each feasible witness using the original groups.
Choose the cheapest completion. The exchange argument proves
optimality, and absence of a feasible witness proves infeasibility.
The candidate enumeration and all arithmetic obey the stated
bound since $r+1\le b$, $b=O((r+k)^2)$, and
$\log(H+1)=O(r+k)$. Exact cost comparisons have polynomial
bit cost in the weight encoding.
\end{proof}

\begin{corollary}[Deterministic actuator optimization]\label{cor:general-control-fpt}
For arbitrary directed state graphs, nonnegative actuator costs,
and all actuators vulnerable, minimum-cost $k$-robust dedicated
placement is solvable deterministically in
$2^{O((d+k)^4)}\operatorname{poly}(n,\ell)$ time.
\end{corollary}
\begin{proof}
Apply Proposition~\ref{prop:control} with $V=E$ and
$\beta_i=k+1$. The rank of the dual matching matroid is $d$,
and its independence oracle is polynomial-time computable by
\eqref{eq:dual-rank} and bipartite matching.
Theorem~\ref{thm:quota-fpt} gives the claim, including the
polynomial-time boundary cases.
\end{proof}

\section{Weighted placement under mixed failures}\label{sec:mixed}
A safe input can satisfy a source group's entire accessibility
requirement. Thus the completion objective is no longer an ordinary
cardinality quota: it may switch between purchasing a safe input
and completing a vulnerable quota. For example, a group with a
safe input costing 5 and two vulnerable inputs costing 4 each,
at $k=1$, has completion costs $5,8,8,8$ for forcing neither,
either, or both vulnerable inputs. The marginal cost of forcing
the second vulnerable input can be zero although the marginal
in an untouched group is three. We use a different method for
this case, exploiting the linear representability of the state
matching matroid.

\subsection{Represented matroids with group survival}
The following algorithm builds on the coefficient method of
Kundu et al.~\cite[Sections~4.3--4.4]{KNS}. Its additional
group state tracks actual actuator purchases, separately from
the small witness used to certify surviving rank.

\begin{theorem}[Represented matroid optimization]\label{thm:linear}
Let a rank-$r$ matroid be represented by columns over a field.
Each element has a nonnegative weight and a safe or vulnerable
status. For any collection of disjoint groups, find a
minimum-weight purchase that remains spanning and meets every
group after any loss of at most $k$ vulnerable purchased elements.
For $r,k\ge1$, this can be solved deterministically in
\[
 2^{O(kr(r+k)\log(r+1))}\operatorname{poly}(n,\ell)
\]
field operations and cost bookkeeping, independent of the number
of groups in the exponential factor. Over $\mathbb Q$ the same
bound holds in bit operations when $\ell$ includes the
representation encoding length. The cases $r=0$ and $k=0$
are polynomial-time solvable.
\end{theorem}
\begin{proof}
The rank-zero case is independent cheapest-group completion.
The zero-failure case reduces to weighted matroid intersection
on the complement, as in Theorem~\ref{thm:quota-fpt}.
Assume $r,k\ge1$, put $b=(k+1)r$, and let $a_e$ be the column
of element $e$. By Lemma~\ref{lem:small}, an optimum purchase
has a resilient witness of size at most $b$. We enumerate
the prescribed final numbers of safe and vulnerable witness
elements with sum at most $b$.

\paragraph{Common future completions.}
Fix a partial witness $W$ of rank $p_0$ and fixed safe and
vulnerable counts. Its remaining witness slots are indexed
$j=1,\ldots,q$, with safe slots first and elements of each
status in processing order. For a fixed future witness
$Z=(z_1,\ldots,z_q)$, use distinct formal tags $T_j$ and formal
$r$-vectors $X_h$, $0\le h<r-p_0+k$. The substitution
\[
 X_h\longmapsto \sum_{j=1}^q u_j T_j^h a_{z_j}
\]
uses independent variables $u_j$. If $J$ is a set of failed
vulnerable future slots, define
\[
 f_J(t)=\prod_{j\in J}(t-T_j)=\sum_{\nu}c_{J,\nu}t^\nu,
 \qquad Y_{J,h}=\sum_{\nu}c_{J,\nu}X_{h+\nu}.
\]
The substitution sends $Y_{J,h}$ to a moment of surviving
columns, with coefficients $u_jf_J(T_j)$.

For a permitted past failure $I\subseteq W\cap V$, choose a
column basis $A_I$ of $W\setminus I$ and put
$\eta=r-\rk(W\setminus I)$. For $|I|+|J|\le k$ form
\[
 D_{I,J}=\det[\,A_I\mid Y_{J,0}\mid\cdots\mid Y_{J,\eta-1}\,],
\]
with value 1 when $\eta=0$. The largest moment index is
$\eta-1+|J|\le r-p_0+k-1$. After substitution, the coefficient
of a product of $\eta$ distinct $u_j$ variables from surviving
slots is the corresponding column minor, multiplied by a
Vandermonde determinant and the nonzero factors $f_J(T_j)$.
Consequently $D_{I,J}$ evaluates to a nonzero polynomial
exactly when the survivors for $(I,J)$ span the matroid.
Distinct slot subsets have distinct monomials, and the tags
are independent indeterminates; over $\mathbb Q$ they may
instead be distinct integers.

Define the product $\Phi_W=\prod_{|I|+|J|\le k}D_{I,J}$,
over permitted past and future failures. The polynomial ring is
an integral domain, so under the substitution for a fixed $Z$,
$\Phi_W$ is nonzero exactly when $W\cup Z$ is resilient.
The substitution is a linear map on coefficient vectors
common to all partial witnesses with the same counts and rank.

\paragraph{Group-by-group dynamic program.}
Process the disjoint groups consecutively, with ungrouped
elements in a separate block. Each element may be skipped,
purchased as a witness element, or purchased for accessibility
alone. Either purchase pays its full cost. Store the partial
witness counts and rank, and for the current group store a
safe-purchase flag and the number of vulnerable purchases,
capped at $k+1$. At the end of each required group accept
only states with the flag set or the count saturated, then
reset the counters. Empty required groups are infeasible.

Within a state, order candidates by purchase cost and retain
a candidate exactly when its coefficient vector for $\Phi_W$
is linearly independent of the previously retained vectors.
If a discarded witness works with some fixed future witness,
its vector is a linear combination of retained vectors of
no greater cost, and the common substitution is nonzero on
at least one of them. That retained witness works with the
same future witness. The group counters and the fact that all
completed groups passed ensure the same future quota-only
purchases also work. At the next processing step, take the
prescribed future decision of an optimum marked purchase
and apply this argument to the generated successor. Induction
over the processing positions preserves an optimum through
all compressions. At the end choose the least costly retained
feasible purchase among the prescribed final count pairs.
The stored witness and group counters certify feasibility.

\paragraph{Complexity.}
There are at most
\[
 N_F=\sum_{i=0}^{\min(k,b)}\binom bi
       \le (eb/k)^k\le (2er)^k
\]
failure factors. The bound by $(eb/k)^k$ follows by evaluating
$(1+x)^b$ at $x=k/b$. A factor has degree at most $r(k+1)$.
There are $O(r(r+k))$ shared variables, including the tags,
because $q\le b$. Thus the coefficient dimension is bounded by
\[
 (1+r(k+1)N_F)^{O(r(r+k))}
   =2^{O(kr(r+k)\log(r+1))}.
\]
Polynomial multiplication and elimination have cost polynomial
in the dimension, the factor count, and the input size;
the determinant expansions fit the bound. The number of
processing positions, target-count pairs and witness states
is polynomial in $n,b,r$, and the current group adds only
$2(k+2)$ possible counter values.

For rational inputs, take a row basis and clear column
denominators. Integer entries have polynomial bit length
in the original input. With tags at most $b$, the logarithm
of the coefficient $\ell_1$-norm of each determinant factor
is polynomial in that bit length and in $r,k$; this follows
from the determinant expansion and submultiplicativity of the
norm. The logarithm of the product norm is at most $N_F$
times that bound. Exact elimination on the coefficient vectors
therefore has polynomial bit cost in their dimension and
coefficient bit lengths. This preserves the stated exponential
factor.
\end{proof}

\subsection{Sampling the dual state matching matroid}
\begin{theorem}[Mixed-failure actuator optimization]\label{thm:random-control}
For a directed state graph with matching deficiency $d$, costs
of total encoding length $\ell$, and arbitrary safe and vulnerable
locations, there is an algorithm running in
\[
 2^{O(kd(d+k)\log(d+1))}
 \operatorname{poly}(n,\ell,\log(1/\delta))
\]
bit operations for $d,k\ge1$ and $0<\delta<1$.
Every returned placement is feasible. On a feasible instance
an optimum is returned with probability at least $1-\delta$;
on an infeasible instance the algorithm never returns a placement.
The cases $d=0$ and $k=0$ are polynomial-time solvable.
\end{theorem}
\begin{proof}
Let $\rho=n-d$ be the maximum state matching size.
Form an $n\times n$ matrix with independent indeterminates
at entries corresponding to bipartite state edges, with rows
indexed by left states and columns by right states. Each basis
of the transversal matroid $\mathcal T$ of size $\rho$ has a
matching to $\rho$ distinct rows, whose determinant is a
nonzero polynomial of degree $\rho$: the matching monomial
cannot cancel with another permutation monomial. Conversely,
any nonzero minor of a numerical realization implies a
matching of its columns. The sampled rank is therefore at
most $\rho$.

Sample all indeterminates independently and uniformly from
$\{1,\ldots,2^a\}$ over $\mathbb Q$, where
\[
 a=n+\lceil\log_2(n/\delta)\rceil .
\]
By the Schwartz--Zippel bound~\cite{Schwartz}, a chosen minor
for any fixed true basis vanishes with probability at most
$\rho/2^a$. There are at most $2^n$ true bases. A union bound
shows that with probability at least $1-\delta$ every true
basis survives. Reject a sample of rank less than $\rho$.

At rank $\rho$, take a basis of the kernel of the sampled
matrix as the rows of a $d\times n$ matrix. Its column matroid
is the dual of the sampled transversal matroid: projection
of the kernel onto a $d$-element coordinate set is bijective
exactly when its complementary $\rho$ columns are independent.
Every sampled primal basis is a true primal basis, so every
sampled dual basis is a true dual basis. Any spanning set
in the sampled dual therefore spans the true dual.

Apply Theorem~\ref{thm:linear} to this representation, using
the source SCCs as the disjoint groups. By
Proposition~\ref{prop:control}, every returned placement is
feasible for the original graph, since all sampled survivor
ranks imply true survivor ranks and the group tests are exact.
On the event that every chosen matching minor survives, the
matroids coincide and the algorithm finds a true optimum
or certifies infeasibility. Otherwise the trial can fail to
return a placement or return a costlier feasible one.

Matching, SCC computation, and kernel elimination take
polynomial bit time. Kernel entries are ratios of minors
with bit lengths $O(n(a+\log(n+1)))$. Substituting rank $d$
into Theorem~\ref{thm:linear} establishes the bound.
For $d=0$, solve the groups independently, and for $k=0$
use weighted matroid intersection on the complement with
source quota one. Both require no sampling.
\end{proof}

The representation sampling above uses a union bound over all
bases for simplicity. For optimization alone, it suffices
to preserve one basis for each survivor of a fixed optimal
small witness; the repository's implementation uses a
smaller sampling bound based on this observation.

\section{Discussion and conclusion}\label{sec:conclusion}
The matching--accessibility decomposition unifies the three
algorithms. For one failure, the flow relaxation handles all
local redundancy constraints, and the common-repair lower
bound controls the cost of restoring global spanning. For
multiple failures, the small witness separates the rank
deficit from the number of source components. With
all-vulnerable locations, completion prices and flat capacities
give deterministic optimization for arbitrary costs. Mixed
failures require a different group decision, which the linear
representation and coefficient-space dynamic program handle
directly.

The approximation guarantee is polynomial in network size
without a small-deficiency assumption. The exact algorithms
are parameterized complexity results: their exponential
factors can be large even for moderate $d$ and $k$. The
bounded-deficiency family in Figure~\ref{fig:running} illustrates
a nontrivial structural regime, but the prevalence of that
regime in measured industrial networks requires separate
empirical study. The model concerns generic structural
controllability under complete input loss; numerical
controllability margins, actuator dynamics and state-coupling
failures are separate design constraints.

The optimal approximation ratio for the one-failure problem
on general digraphs remains open. Redundancy deletion and
backups shared by larger sets of critical inputs are possible
directions for improving practical augmentation. For mixed
failures, a deterministic algorithm with parameter dependence
only on $d+k$ and polynomial dependence on $n$ remains a
target for the represented-matroid route. The complexity of
parameterization by $d$ alone also merits investigation.

\paragraph{Computational verification.}
The reference implementations and finite-instance audits
compare the approximation and exact algorithms with
independent exhaustive computations on small matroids and
state graphs. The tests exercise closure saturation, quota
completion, mixed-status purchases, and sampled representations.
The mathematical guarantees follow from the proofs above;
the audits check the corresponding implementations.

\paragraph{Code availability.}
The source, algorithms, and verification scripts are available
at \url{https://github.com/dongxuelian2/robust-structural-controllability}.
The paper guide lists reproducible commands and proof notes,
including the earlier stand-alone matroid results.

\paragraph{Acknowledgment.}
We thank Saket Saurabh for suggestions concerning the group
counters and rank-checked sampling in the mixed-failure algorithm.

\paragraph{AI assistance.}
Language-model tools from OpenAI, Google, and Anthropic were used
in developing and checking arguments, preparing verification
code, and editing the manuscript. The authors are responsible
for the mathematical statements, code, and final text.

\end{document}